# Nonlinear Quantum Theory, Development of the Superposition Principle and Possible Tests

Yi-Fang Chang

Department of Physics, Yunnan University, Kunming 650091, China

(E-mail: yifangchang1030@hotmail.com)

**Abstract**: First, we point out that the present applied superposition principle is linear, it must be developed into a generality. Next, the linear operators and equations should be developed nonlinearly. We suppose $p_\mu = -i\hbar(F\frac{\partial}{\partial x_\mu} + i\Gamma_\mu)$, so Klein-Gordon equation and Dirac equations turn out to be $(F^2 \Box + \Gamma_\mu^2 - m^2)\varphi = -J$ and $\gamma_\mu(F\partial_\mu + i\Gamma_\mu)\psi + \mu\psi = j$, respectively. The corresponding Heisenberg equation is $\frac{dL}{dt} = \frac{1}{ihF_0}(LH - HL) + Q$. The quantum commutation and anticommutation belong to $F$ and $\Gamma_\mu$. This theory may include the renormalization, which is the correction of Feynman rules of curved closed loops. We think the interaction equations are nonlinear. Many theories, models and phenomena are all nonlinear, for instance, soliton, nonabelian gauge field, and the bag model, etc. The superluminal entangled state, which relates the nonlocal quantum teleportation and nonlinearity, should be a new fifth interaction. Moreover, the NL effects exist possibly for four interactions, for single particle, for high energy, and for small space-time, etc. The relations among NL theory and electroweak unified theory, and QCD, and CP nonconservation, etc., are expounded. Finally, some known and possible tests are discussed. The NL theory relates the possible decrease of entropy in isolated system.

**Key Words**: quantum theory, nonlinearity, superposition principle, interaction, equation, operator, model, test.

**PACS**: 03.65.Ta; 05.45.-a; 03.65.-w; 03.65.Pm; 12.15.-y; 12.10.Dm

## 1.Introduction

The quantum mechanics is very beauty theory, and obtained great success. But it encountered also many difficulties, therefore physicists are trying to develop the quantum theory. From the de Broglie-Bohm nonlinear wave mechanics [1,2] and the Heisenberg spinor unified field theory [3,4] until present [5-16], the developments of various nonlinear (NL) theories are always remarkable. Thacker discussed an exact integrability in quantum field theory and statistical systems, which include the nonlinear Schrodinger model and equation [5]. Gerjuoy, et al., researched a unified formulation of the construction of variational principles, whose equations may involve difference or differential or integral operators, while they may be homogeneous or inhomogeneous, linear or nonlinear, self-adjoint or not, and may or may not represent time-reversible systems [6]. Kivshar, et al., discussed dynamics of solitons for nearly integrable systems, and focused on four classical nonlinear equations: the Korteweg-de Vries, nonlinear Schr?dinger, sine-Gordon, and Landau-Lifshitz equations [7].

Recent Marklund, et al., considered strong-field effects in laboratory and astrophysical plasmas and high intensity laser and cavity systems, which relates to quantum electrodynamical (QED) nonlinear photon-photon scattering, and nonlinear collective effects in photon-photon and photon-plasma interactions [8]. Zhang, et al., demonstrated temporal and spatial interference between four-wave mixing (FWM) and six-wave mixing (SWM) channels of these two nonlinear optical processes in a four-level atomic system [9]. Paternostro, et al., shown violations of Bell inequality for Gaussian states with homodyne detection and nonlinear interactions using nonlinear unitary operations, and this approach may be extended to other applications such as entanglement distillation where local operations are necessary elements besides quantum entanglement [10]. Falco, et al., illustrated relation between the nonlinear Schrodinger-Langevin-Kostin equation and quantum annealing [11]. The quantum self-trapping phenomenon of a dipolar Bose-Einstein condensate (BEC) as a nonlinear effect and symmetry breaking are studied by Xiong, et al. [12]. Vukics, et al., searched cavity nonlinear optics with few photons and ultracold quantum particles, and the quantum character of particle motion generates nonlinear field dynamics, and derived a corresponding effective field Hamiltonian containing all the powers of the photon number operator, which predicts nonlinear phase shifts. Moreover, simulations of the full particle-field dynamics show significant particle-field entanglement [13]. Lu, et al., studied the nonlinear dynamics of collective excitation in Heisenberg spin chains through the magic angle, in which two bright and dark controlling solitons of nonlinear excitations appear [14]. Buric studied the nonlinear dissipative classical or quantum system, and dynamics of the interqubits entanglement in which the semiclassical oscillator is chaotic rather than regular [15]. Pezze, et al., researched the Heisenberg limit and the creation of useful entangled states by the nonlinear dynamical evolution of two decoupled Bose-Einstein condensates or trapped ions [16].

But, so far various NL theories as by other methods of modified quantum theories are not the major current of the high physics. At another aspect, the NL theories have made great progress in many ranges of mathematics and physics [17-19], in particular, the soliton and chaos. Hanggi, et al., discussed reaction-rate theory in fifty years after Kramers, which relates nonlinear science of physics, chemistry, engineering, and biology [20].

We studied the logical structure of the quantum mechanics, and proposed some approaches to overcome the cardinal trouble of the quantum theory [21]. Bohr, Heisenberg [22], Dirac and Yukawa, et al., thought that some basic concepts must are developed. But, what is the development? It is troubled. Based on some very essential concepts of mathematics and physics, we propose a possible NL quantum theory. In sec.2 the superposition principle is analyzed. We think that it should be developed to the more general NL superposition principle [21,23]. In sec.3 the present linear operators are developed, so we obtain some new equations, which may be NL. In sec.4 the quantization rules are approached. In sec.5 we try to make a research for the essence of the perturbation theories on higher order and of the renormalization. In set.6 and 7 we think that the NL effects exist possibly for four interactions, for single particle, for high energy, and for small space-time, etc. The relations among NL theory and electroweak unified theory, and QCD, and CP nonconservation, etc., expounded. Finally, some known and possible experimental tests for NL theory are discussed.

## 2. Development of the superposition principle

Feynman had pointed out [24]: "In analysing physical phenomena we assume the validity of two principles: 1) special relativity, 2) the superposition of quantum mechanical amplitudes. As far as we know the principles are exact. The stage upon which all is played is the flat Minkowski space." Indeed, the present quantum mechanics is based on the superposition principle (SP). It is very clear that this connection expounded by Dirac [25]. Therefore this seems to hold: "The superposition principle of quantum theory is an independent hypothesis", "it is independent of whether such equations are linear——or whether field equations exist!" [26].

Notably, this is the general SP as Dirac, et al., had expounded. It holds always! But in the concrete cases the present applied SP is only special linear, i.e., $\psi = \sum_n C_n \psi_n$. Therefore, although various expressions of SP are different, but we require only: "It follows from the principle of superposition of states that all equations satisfied by wave functions must be linear in $\psi$" [27,28,25]. In fact many equations of wave functions $\psi$ (and $\varphi$, $A_\mu$) already were NL, for example, for some interactions. Their solutions (i.e., the wave functions) did not agree with the linear SP, in particular, when this solution is a soliton. It shows that the superposition principle must be developed.

So long as "between these states there exist peculiar relationships", "state may be considered as the result of a superposition of two or more other states, and indeed in an infinite number of ways" [25]. This general SP includes the NL superposition principle——Backlund transformation of soliton [17]. Furthermore, in the nonlinear quantum theory the equations and operators are nonlinear [29], so the present applied linear superposition principle should be developed. If the linear SP connects with the probability interpretation of wave, this interpretation will be different for the quantum field theory.

## 3. Extension of linear operators and NL equations

The present quantum theories postulate that "all of operators possess the property of linear" [30]. It is that $p_\mu = -i\hbar \frac{\partial}{\partial x_\mu}$, etc., hold always, which are based on the linear SP, or on Fourier transform, or on the association of particles with plane wave. But we know that the linear SP and Fourier integrals of NL waves had not held from their mathematical conclusion [17-19]. Therefore, we think the waves of particles are NL waves in some cases, for instance, particles correspond to soliton-waves, so the operators may not be linear.

The NL theories are very complex, and the NL approach also has many schemes. Here we point out some possible methods:

$$\text{A.} \quad p_\mu = -i\hbar \frac{\partial}{\partial x_\mu} \Rightarrow -i\hbar (\frac{\partial}{\partial x_\mu} + i\Gamma_\mu). \tag{1}$$

If the plane wave is the generalized $\psi(x,t) = u(x,t)e^{ip_\mu x_\mu / \hbar}$, which may be the NL wave, there will be $\frac{\partial \psi}{\partial x_\mu} = \frac{\partial u}{\partial x_\mu} e^{ip_\mu x_\mu / \hbar} + \frac{i}{\hbar} p_\mu \psi$, and $p_\mu = -i\hbar(\frac{\partial}{\partial x_\mu} - \frac{1}{u}\frac{\partial u}{\partial x_\mu})$. Such Eq.(1) is analogue with $P_\mu = p_\mu - \frac{e}{c}A_\mu$ of electromagnetic field, with $\nabla_\mu = \partial_\mu - \Gamma(A)$ in the gauge theory, with the covariant derivative.

$$\text{B.} \quad p_\mu = -i\hbar \frac{\partial}{\partial x_\mu} \Rightarrow -i\hbar F \frac{\partial}{\partial x_\mu}. \tag{2}$$

It is analogue with the NL curved space-time, and with the general theory of relativity of the gravitational field.

C. The statistical approach.

A and B can be combined by

$$p_\mu = -i\hbar (F \frac{\partial}{\partial x_\mu} + i\Gamma_\mu). \tag{3}$$

It is a Hermitian operator still. $\Gamma_\mu$ and F are additive term and corrected factor, respectively. Here $\Gamma_\mu$ and F may be NL. When $\Gamma_\mu \to 0$ and F→1, the operator and theory return to an original forms.

In this case the generalized Klein-Gordon equation is:

$$|F \frac{\partial}{\partial x_\mu} + i\Gamma_\mu|^2 \varphi - m^2 \varphi = (F^2 \Box + \Gamma_\mu^2 - m^2)\varphi = -J. \tag{4}$$

The Dirac equations are:

$$\gamma_\mu (F \frac{\partial}{\partial x_\mu} + i\Gamma_\mu)\psi + \mu\psi = j. \tag{5}$$

Let J=0, the Schrodinger equation is:

$$i\hbar F_0^2 \frac{\partial}{\partial t}\psi' = -\frac{\hbar^2}{2m}F_r^2 \nabla \psi' + (1 - F_0^2)\frac{1}{2}mc^2\psi' - \frac{\hbar^2}{2m}\Gamma_\mu^2 \psi', \tag{6}$$

which $\psi' = \varphi e^{imc^2 t/\hbar}$. If $\Gamma_\mu = c\psi'$, Eq.(6) will be a cubic Schrodinger equation. When F=1, Eq.(6) becomes to $-\frac{\hbar^2}{2m}\nabla \psi' = (E + \frac{\hbar^2}{2m}\Gamma_\mu^2)\psi'$, which is the same with the Schrodinger equation of the general theory of relativity [31]. Here the potential is $U = -\frac{\hbar^2}{2m}\Gamma_\mu^2$.

If F=1, this theory will include various gauge theories, in which the field equations of nonabelian group are NL, and will include the additive equivalent mass terms which are derived by interaction fields.

The new F and $\Gamma$ are probably dependent on the types of particle and of interaction, on energy and space-time, etc. Let them be various functions and higher derivatives, we will derive various equations, which include the mathematical NL equations. We may correct equivalently the Lagrangian density, and derived the corresponding quantum equations [21].

Because

$$H\psi = i\hbar(F_0 \frac{\partial}{\partial t} + i\Gamma_0)\psi, \tag{7}$$

so the operator U(t), which is played on the transformation from a Schrodinger picture into a Heisenberg picture, satisfies the same equation. When H, $F_0$ and $\Gamma_0$ are not a function of time, the solution is

$$U(t) = \exp[-i(H + \hbar\Gamma F_0)t/\hbar F_0]. \tag{8}$$

The equation of motion is

$$\frac{dL}{dt} = \frac{1}{i\hbar F_0}(LH - HL) + Q. \tag{9}$$

If $\Gamma_0$ is real or imaginary quantity, so

$$Q = \frac{1}{iF_0}(L\Gamma_0 \mp \Gamma_0 L) \tag{10}$$

will be commuting or anticommuting operators.

Heisenberg equation connects with the Poisson bracket, and with Liouville equation. They correspond to the Hamilton canonical equations, but which are applicable only for holonomic conservative system in classical mechanics, where $p_i$ and $q_i$ are symmetrical. The corresponding Lagrangian equation

$$\frac{\partial L}{\partial \psi_\alpha} - \frac{\partial}{\partial x_\mu}\frac{\partial L}{\partial \partial_\mu \psi_\alpha} = 0, \tag{11}$$

hold also only for this system. For the nonconservative system the symmetry is broken,

$$\frac{dq_i}{dt} = [q_i, H], \quad \frac{dp_i}{dt} = [p_i, H] + \overline{Q_i}. \tag{12}$$

For the nonholonomic system the equations are obtained by Routh, et al. For interactions, esp., the self-interaction and virtual process, the system may not be the holonomic conservative, so Heisenberg equation should be extended to

$$\frac{dL}{dt} = [L, H] + Q. \tag{13}$$

This is analogue with the extension of classical mechanics and of the Liouville equation, which adds a term $S_q$. When $Q = [L, \Gamma_0]_\pm + J'$, Eq.(13) is an extension of Eq.(9). In the present Heisenberg equation H should be Hamiltonian $H_0$ of the holonomic conservative system, i.e., $dL/dt = [L, H_0]$. It just corresponds to the interaction picture.

## 4. Quantization rules

In quantum theory bosons and fermions are commuting and anticommuting relations, respectively. In our nonlinear theory the quantized conditions (commuting relations) are:

$$x_\mu p_\nu - p_\nu x_\mu = i\hbar F \delta_{\mu\nu} - i\hbar (x_\mu F - F x_\mu)\frac{\partial}{\partial x_\nu} + \hbar(x_\mu \Gamma_\nu - \Gamma_\nu x_\mu). \quad (14)$$

If $F$ and $\Gamma_\mu$ are commuting with $\frac{\partial}{\partial x}$, and $F$ is commuting with $\Gamma_\mu$, so there will be

$$p_\mu p_\nu - p_\nu p_\mu = \hbar^2 (\Gamma_\mu \Gamma_\nu - \Gamma_\nu \Gamma_\mu). \quad (15)$$

The commution of $p$ operators reverts to the commution of $\Gamma$. If F and $\Gamma$ are anticommuting with $x_\mu$, Eq.(14) will become

$$x_\mu p_\nu + p_\nu x_\mu = -i\hbar F \delta_{\mu\nu}. \quad (16)$$

Conversely, if F and $\Gamma$ commute with $x_\mu$, Eq.(14) will be only more a factor F than the original relations, and the anticommuting relations turn out commuting. Therefore, this different property of bosons and fermions belongs to the different characters of F and $\Gamma$. This will be able to combine the supersymmetry [32-34].

The general anticommuting relations are

$$x_\mu p_\nu + p_\nu x_\mu = i\hbar F \delta_{\mu\nu} - i\hbar(x_\mu F \frac{\partial}{\partial x_\nu} + F \frac{\partial}{\partial x_\nu} x_\mu) + \hbar(x_\mu \Gamma_\nu + \Gamma_\nu x_\mu). \quad (17)$$

If the definition of corresponding annihilation and creation operators is invariant [35], there will be $\{a_{\alpha'}, a_{\alpha'}^+\} = F + A$. The number operator is $N_{\alpha'} = a_{\alpha'}^+ a_{\alpha'}$, then

$$N_{\alpha'}^2 = a_{\alpha'}^+ a_{\alpha'} a_{\alpha'}^+ a_{\alpha'} = a_{\alpha'}^+ (F + A - a_{\alpha'}^+ a_{\alpha'}) = a_{\alpha'}^+ a_{\alpha'}(F + A) = N_{\alpha'}(F + A), \quad (18)$$

so $N_{\alpha'}$=0 and F+A. When F=1 and A=0 (or A<<0), the eigenvalues are 0 and 1. The low energy is so, the Dirac hole theory hold still, and this case obeys the Pauli exclusion principle. If the above conditions do not hold, so the Pauli exclusion principle may not hold. It derives the conclusion expounded by Santilli, et al. [36], and by me from the NL theory, etc. [37-40,21,23], and by Greenberg, et al. [41,42]. The conditions for the existence of this conclusion are the same.

If F and $\Gamma_\nu$ are anticommuting with $x_\mu$, so Eq.(17) is also only more a factor F. Let

$$a = \frac{1}{\sqrt{2F}}(q + ip), \quad a^+ = \frac{1}{\sqrt{2F}}(q - ip), \quad (19)$$

so $[a, a^+] = 1$, and various conclusions of quantum field theory hold still in this case.

Furthermore, when Bose-Einstein and Fermi-Dirac statistics are unified at high energy [37], the quantized condition also must be developed. The extension of the quantization rules corresponds to the development of Lie algebra [43]. If both are combined, the quantum theory will be able to be developed better, I think.

### 5. Perturbation, renormalization and Feynman rules

Through careful analysis for quantum electrodynamics [21], we think that the perturbation theories above two order connected already with NL problems, because higher-order of $\varphi$, $\psi$ and $\overline{\psi}$ appear in a term here. The perturbation theory tries to describe asymptotically NL phenomena by the linear method. The corrections of different order of the perturbation theory are various NL corrections. For instance, the radiative corrections are self-interactions, which are just NL. Conversely, according to the present theory the NL photon-photon interaction [8,13] can be only self-energy of photon which is also the higher perturbation. Further, $H_1 H_2$ of the perturbation theory is analogue with the NL Lagrangian $L_1 = \alpha(E^2 - H^2)^2 + \beta(EH)^2 + ...$ in the Born NL electrodynamics.

The perturbation theory holds, which shows that the NL effects are smaller. If theories are essentially NL, so the perturbation theory may not hold, for example, for weak interactions. Only in the electroweak theory the NL Higgs fields of breaking symmetry are not considered, the theory may be renormalizable.

If Feynman rules are directly applied to calculate diagrams containing closed loops, the corresponding integrals will be divergence at large momentum. For this reason the theories must be renormalizable. We think that it is equivalent to the correction of Feynman rules of closed loops. In Feynman diagrams the closed loops of internal lines are curved, and their figures are curved geometry, which is NL. Indeed the closed loops correspond to the NL virtual processes. The external lines may always be straight lines. Therefore, the present Feynman rules should correspond only to straight lines, while Feynman rules of curved lines must consider the modified effects of curved geometry, and must be developed. This is namely the renormalization. It can belong to correct the propagators S and $\Delta$, which are the internal lines of bosons and fermions, and correct the vertex functions $\Gamma$, which correspond to interactions, and belong to add masses. $\Gamma_0 \to Z_1^{-1}\Gamma$, $S_0 \to Z_2 S$, $\Delta_0 \to Z_3 \Delta$, they are the correction of Feynman rules [44]. Moreover, the propagators and the vertices may just form the closed loops of Feynman diagrams. In the Riemannian geometry the straight line $du^i = 0$ is developed to the geodesic line

$$du^i (1 + \Gamma_{kl}^i du^k \frac{dx^l}{dx^i}) = 0, \qquad (20)$$

whose form is analogue with the propagator $\Delta' = Z_3 \Delta = \Delta(1 + G^2 \Pi'(-\mu^2))$ of the renormalization. The divergences of the present linear theories are analogue with Olbers divergence in the classical astronomy of flat space-time, too.

The Lagrangian and equations of the renormalization are consistent with Eqs.(4) and (5). F corresponds to the renormalization of wave functions and of parameters, and $\Gamma_\mu$ corresponds to the additional mass of the renormalization. For the scalar fields, $\Gamma$, S and $\Delta$ are Green functions of system of coupling fields. For the fermion fields of renormalization,

$$(\gamma^\mu \partial_\mu + \mu) Z_2 S = [\gamma^\mu (Z_2 \partial_\mu) + (Z_2 \mu)] S , \quad F = Z_2, \quad i\gamma^\mu \Gamma_\mu = (Z_2 - 1)\mu - \delta\mu. \quad (21)$$

Such the corrections of equations are just equivalent to the renormalization. For the bosons fields, $F^2 = Z_3$, $\Gamma_\mu^2 = \delta\mu^2 - (Z_3 - 1)m^2$.

The renormalization of gauge theories is equivalent to the insertion in the Lagrangian of local counteracting terms, and is equivalent to the renormalization of the parameters involved in the Lagrangian. Eqs.(5) are analogue with the generalized Callan-Symanzik equations [45]:

$$[\sum_{i=1}^n \beta_i(x) \frac{\partial}{\partial x_i} + \gamma(x)] \Gamma_0(x) = \Gamma_1(x). \quad (22)$$

These all are based on the supposition that the present form of quantum field theory is exact, only it is corrected to a NL form here.

If j=0 and $\Gamma_\mu$ =0, Eqs.(5) will become to

$$\gamma_\mu \frac{\partial}{\partial x_\mu} \psi + \frac{mc}{\hbar F} \psi = 0, \quad (23)$$

which has only a different mass. Therefore, the magnetic moment of electron should be $-\frac{e\hbar}{2mc} F\sigma'$.

**6. Various phenomena, interactions and nonlinear theory**

In the quantum mechanics there is an old question: The wave packet, which is obtained by the superposition of the plane waves, will spread necessarily. It corresponds to the linear theory [1]. Therefore the stability of single particle cannot be described by the present wave property. But the NL solitary waves keep stability, hence Lee, et al. [46], supposed particles are solitons. We think that when the wave-particle duality holds, particle may only be soliton. This is development of the duality. The elastic collisions are analogue with the collisions of solitons. The fractional charge may be obtained in the soliton theory.

The wave functions of particles are usually

$$\psi(x,t) = \psi(x,0) e^{-iEt/\hbar}. \quad (24)$$

Because the energy of particle is real, the probability $|\psi(t)|^2 = |\psi(0)|^2$ of discovered particle is not function of time, therefore this cannot describe change of particle. In order to describe decay, resonance and binding state, etc., of particles, energy must be extended. The excited state has a complex energy $E = E_0 - (i\Gamma/2)$ [47], so $|\psi(t)|^2 = |\psi(0)|^2 e^{-\Gamma t/\hbar}$, i.e., the exponential decay. The corresponding wave functions are

$$\psi(t) = \psi(0) \exp[-(iE_0 t/\hbar) - (\Gamma t/2\hbar)]. \quad (25)$$

Therefore on the decay and change, etc., of particles, no matter what interactions they are, the plane wave must be extended. The operator of $E_0$ is

$$E_0 = i\hbar(\frac{\partial}{\partial t} + \frac{\Gamma}{2\hbar}), \qquad (26)$$

which belongs to a simple Eq.(1).

On the interchange, annihilation, creation and decay of particles, they are already some NL effects in classical electromagnetic theory, i.e., the Born-Infeld NL electrodynamics, and so are they in the quantum theory.

We discuss four interactions. Gravitational interactions are the known NL general relativity. Strong and weak interactions all must pass the NL Higgs field, then masses can only be obtained, whatever they are represented by Weinberg-Salam electroweak theory, or by QCD, or by the grand unification theory, or by the superstring. Furthermore, strong, weak and gravitational interactions all are NL, too. Even electromagnetic interactions are also NL in the strong field and high energy, esp., for photon-photon interactions. The experiments proved that photons at high energy ($\geq$ several GeV) shown the similar properties of hadrons [48,49].

The present theories suppose that four interactions all are the gauge fields. The geometrical interpretation of Yang-Mills field is that fields are analogue with the tensor fields [50]. An identity which is the analogy of the Bianchi identity in the theory of gravity is obtained, and the fields determine the curvature of this space, For the nonabelian gauge group $f^{abc} \neq 0$, the Lagrangeian of the YM fields has the NL terms of higher order than two, i.e., these fields possess the self-interactions. It proves that strong and weak interactions, which have SU(3) and SU(2) symmetry, and must be NL theories. When they are developed by analogy with the theory of gravity, this theory connects with the strong-gravitation unification theory [51,52], and with the supersymmetry theory [32-34].

Further, the fundamental field equations must be nonlinear in order to represent interaction [3]. When we solve a set of coupling equations of interactions, the NL equation must be obtained. In the non-relativity quantum mechanics, which is the most successful part in quantum theory, various interactions all are became the corresponding potential. But this method does not hold for high energy and relativity. For spinor fields the Heisenberg unified equation is NL [3], namely, F=1, $\mu$=0 and $\Gamma_\mu = -i\gamma_\mu l_0^2 \overline{\psi}\psi$ in Eqs.(5). For scalar fields the Higgs field and the Goldstone equation are NL in symmetry breaking. For vector fields the equations of nonabelian group are NL. Only the nonabelian gauge fields possess NL self-interactions, and the asymptotic free property of four space-time [53]. Therefore these asymptotic free experiments required that theories must be NL.

The self-energy and self-interaction must be NL, as Heisenberg [3,4] and Burt [26] pointed out. Even at the classical level virtual process, vacuum excitation and fluctuation all are NL. For instance, Bohm thought that random fluctuations correspond to NL.

Further, gravitational interactions are NL. All of body is in a gravitational ground, which is usually very weak, and which may be neglected, esp., for two order NL effects. But at high energy various interactions tend to the unification [37,21], so the NL gravitational interactions cannot be neglected, various NL effects should appear.

According to the special relativity the velocity of light in the vacuum is a limit velocity, so any velocity cannot be superluminal. We proposed that the basic principles of the special relativity should be redefined more suitably as [21,54,55]: I. The special relativity principle, which derives necessarily an invariant velocity $c_h$. II. Suppose that the invariant velocity $c_h$ in the theory is the velocity of light in the vacuum. The extensive special relativity has only a principle I, whose formulations are invariant only c→$c_h$. Such the superluminal is possible. Recently, various superluminal phenomena are observed [56-62]. In particular, Wang, et al., measured a very large superluminal group velocity that exceeds about 310 times faster than the speed of light in a vacuum [60]. Salart, et al., tested the speed of spooky action at a distance [62]. This is a speed of the entangled state, which relates the nonlocal quantum teleportation [63,64], and violates Bell inequalities.

At present there are only four interactions: gravitational and electromagnetic fields whose action distances are infinite (i.e., long-range forces), strong and weak fields whose action distances are very finite (i.e., short-range forces). New researches shown that the entangled state should be is a new fifth interaction [54,65]. Its action distance is middle-rang, i.e., neither infinite nor very short, and its strength is also middle one. It should connect with nonlinearity. This seems to like the thought field [66].

The five interactions may be exhibited completely in Fig.1. The entangled state is in a middle state among the known four interactions. It seems to be fully enlightening guidance.

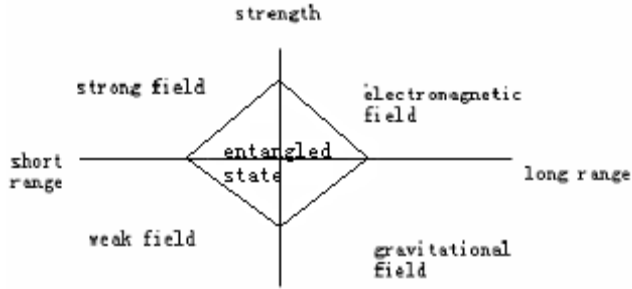

Fig.1 Relation among five interactions

The entangled state may be the following representation:

$$X_{00} = \frac{1}{\sqrt{2}}[|\frac{1}{2}>_A|-\frac{1}{2}>_B - |-\frac{1}{2}>_A|\frac{1}{2}>_B], \quad (27)$$

$$X_{10} = \frac{1}{\sqrt{2}}[|\frac{1}{2}>_A|-\frac{1}{2}>_B + |-\frac{1}{2}>_A|\frac{1}{2}>_B]. \quad (28)$$

**7. Some nonlinear models and equations**

At high energy and high excitation particles change easily. High energies correspond to small space-time. Small space-time connects with the NL theory, and derives cut-factor $l_0$, which is equivalent to the non-local theory, and then divergence is avoided. It is consistent that de Broglie thought only small space-time must be corrected [1]. This connects with the composition of particles, and with quark-confinement, etc. Let $\Gamma_\mu = gr$ in Eq.(25), so the potential U→0 is asymptotic free when r→0; and is quark confinement since U is proportional to $r^2$. $U = -\frac{\hbar^2}{2m}g^2r^2$ is also the harmonic oscillator potential, and may derive the Regge mass formula. Therefore the NL QCD holds better only for high energies and small range at present.

For single particle, or for short-range strong and weak interactions, which relate to few particles, or for small space-time, the probability-waves all are meaningless [21]. Therefore, the essence of wave-property and the significance of wave function all must be reunderstood. Although the approximation of solitary waves may be became the plane wave, as NL equations all may become approximately linear equations. But for larger space-time, for lower energy, for usual electromagnetic interaction and for free particle which does not consider a self-energy, etc., when the NL effects do not exist, or may be neglected, then the present theories can only hold very good. Only the electron and muon, etc., which are few point particles, as well as the long-range electromagnetic interactions, which correspond to the Abelian group, can be described exactly by the linear theory. This is just QED.

The Lagrangian of electromagnetic field is

$$L = -\frac{1}{4}F_{\mu\nu}F_{\mu\nu}. \tag{29}$$

Here $F_{\mu\nu} = \partial_\mu A_\nu - \partial_\nu A_\mu$. Assume that $\partial_\mu A_\nu \Rightarrow (\partial_\mu - i\varepsilon A_\mu)A_\nu$, i.e., F=0 and $\Gamma_\mu = -\varepsilon A_\mu$, so L is the Lagrangian of Yang-Mills field of nonabelian group in vacuum. For spinor fields assume that

$$\partial_\mu \Rightarrow \partial_\mu - \frac{1}{2}ig\lambda^a A_\mu^a, \tag{30}$$

so we derive QCD from QED. In QCD there has a pseudoparticle solution (instanton) [67,68], which shows that strong interactions must be the NL equations. Weak interaction fields can be obtained by a similar way. It combines with the electromagnetic fields, and then the Weinberg-Salam electroweak theory may be derived. We must add the NL Higgs fields, and then particles can only obtain masses. Let F=1,

$$\Gamma_\mu^2 = \lambda^2\varphi^2 - (g\tau^a A_\mu^a + g_1 B_\mu)^2, \tag{31}$$

and $J = G\overline{L}R$, so Eq.(4) becomes the equation of Higgs field in the W-S theory.

In order to agree with the Yukawa interactions and gauge fields, we must introduce into gluons in quark model. Gluons may possess a self-interaction, and may form a gluon-ball, so the equation of gluon must be NL.

The SLAC bag model is based on the soliton-solution of NL $\varphi$-field [69]. Inversely, Friedberg and T.D.Lee thought that particles are solitons, so the bag model is obtained [70]. The equations of bootstrap are NL. The equations of superfield are mainly NL, too.

In all interactions which include neutrino, the parity nonconservation is usually belonged to the helicity of neutrino, namely, the factor $(1 \pm \gamma_5)$ is introduced. Here the corresponding corrected factor F may be $(1 \pm \gamma_5)$. Further, if F possesses some properties, for example, the parity acts on some particles and antiparticles for weak interactions, and F is different, the parity nonconservation of all weak interactions will be able to be derived. This seems to prove also that weak interactions are NL.

$K^0$ mesons supply a very good example on the linear superposition, but it is discovered unexpectedly that the CP of $K^0$ mesons is nonconservation. This has not still a recognized interpretation at present, one of better theories is the $K^0$-$\overline{K}^0$ mixture and the Wolfenstein very weak interaction theory [71]. We think that the CP nonconservation is probably a result of the small NL superposition of $K^0$-$\overline{K}^0$. Further, when the masses of particles increase, for instance, for $K^{*0}, D^0, B^0$, etc., which may be similar superposition, the NL effects are probably larger. Lee proved the CP nonconservation is derived when we interchanged the NL Higgs particles [72].

**8. Possible experimental tests**

1989 Weinberg proposed a nonlinear generalization of quantum mechanics (NLQM) and precision tests of quantum mechanics [29]. A hyperfine transition in the ground state of $^9Be^+$ was used to test NLQM. Bollinger, et al., searched for a dependence of the frequency of a coherent superposition of two hyperfine states on the populations of the states, and are able to set a limit of $4 \times 10^{-27}$ on the fraction of binding energy per nucleon of the $^9Be^+$ nucleus that could be due to nonlinear corrections to quantum mechanics [73]. Gisin researched NLQM and the superluminal communications [74]. Chupp and Hoare observed coherence among the four magnetic sublevels of freely precessing $^{21}Ne$, and a test of linearity of quantum mechanics. Nonlinear corrections to quantum mechanics are found to be less than $1.6 \times 10^{-26}$ of the binding energy per nucleon of $^{21}Ne$ [75]. Walsworth and Silvera extended NLQM to systems of composite, or multivalued, spin, such as atoms and molecules [76]. This can exhibit observable nonlinear behavior in a composite spin system.

We think that some known experiments shown the existences of the NL effects. In example, photon-photon interacts in electromagnetic fields; the strong and weak interactions possess symmetry of the nonabelian group; particles regard as the solitons or solitary waves in some experiments, etc. Further, the possible tests are:

A. Direct test. 1.Under some conditions, for instance, single particle, strong or weak interactions, high energies, small space-time range, etc., the NL wave or the NL interaction will appear probably, and the NL superposition principle will agree with quantum phenomena. For the linear superposition the most direct test is the interference and diffraction of particles on the crystal. But, so far there has not a crystal whose space is $10^{-13}$ cm on the Earth, so the same experiments will not be able to be realized. In the coherent processes measured accurately, the NL effects should appear. If the NL superposition exists, the experiments will have deviation from results of the linear superposition. For instance, when we consider the high order perturbation and the photon-photon interactions, etc., the experiments of interference and of diffraction in electromagnetic interaction possess probably the small deviation from the present theory.

In the nucleus the wave function and properties of single nucleon (p, n) may determine by Dirac equations and by experiments, according to the present linear superposition principle, the theories compare with the experiments, we shall be able to determine that whether the linear superposition principle holds for strong interaction or not. The experiments of $\mu$-$Fe^{58}$ collision show that the valence quarks in iron-nuclei carry less momentum than they should carry in an isolated proton, and the ocean quarks are in greater number (EMC effect) [77,78]. It seems to imply that the linear superposition principle is different for the heavy nuclei.

2. The particles possess some characters of NL theory and NL wave [18], for instance, soliton, fractal and chaos, the recurrence of system under the periodic boundary condition, and the self-focusing of waves in the NL fields.

The recurrence seems to correspond to the interchange of $K^0$ and $\overline{K}^0$ by weak interaction, i.e., $K^0 \leftrightarrow 2\pi \leftrightarrow \overline{K}^0$, and to the oscillation theory of neutrino. In the nucleus although the interchange between neutron and proton is very quick, but the structure of stationary nucleus and the isotopic abundance ratio possess macroscopic invariant. This seems to be a recurrence. The vacuum fluctuation and various virtual processes are probably recurrence, too.

B. Indirect test. 1.Various NL equations and their solutions, for example, Higgs equation, NL $\sigma$ model and soliton, instanton, etc., which may describe some experiments. 2.The present theories do not agree with the experiments, in which the many can be explained by the NL theories, for example, the CP nonconservation. But it is only one of possible explanation.

Moreover, Valentini discussed the violation of wave-particle complementarity in NLQM [79]. It is shown, in the context of a double-slit experiment, that small nonlinearities of the Schr?dinger equation, considered by Weinberg [29], will in general allow a simultaneous measurement of particle paths and interference, unless the nonlinearities are of a special form such that nearby initial states remain so at all later times. Polchinski shown that Weinberg NLQM leads either to communication via Einstein-Podolsky-Rosen correlations, or to communications between branches of the wave function [80].

Peres proposed that the nonlinear variants of Schrodinger equation violate the second law of thermodynamics [81], which is a comment on the letter [5]. We discussed possible decrease of entropy in isolated system [82-88], and whose necessary condition is existence of internal interactions. This possibility relates especially the NL theory. A sufficient and necessary condition of decrease of entropy is discussed quantitatively [88]. Further, we calculate quantitatively decrease of entropy in an internal condensed process [84] and in the ordering phenomena and nucleation of thermodynamics of microstructure [88]. Some possible tests are discussed [82-88].

In a word, the NL theory is very difficult and complex. At present our approach possesses some indeterminacy. This is a weakness, but it opens up a broad way for developments of theories. Furthermore, the entire domain of science is passing almost through a direction from linear to NL. Our approach is different with many other NL quantum methods, it is based on the earliest possible starting point——the development of the linear superposition principle and the generalized NL operators.